\documentstyle[12pt,epsfig]{ioplppt}  

\begin{document}
\jl{8}                                  

\title {Computing Lyapunov spectra with continuous Gram-Schmidt 
orthonormalization}

\author{F Christiansen\dag\ and H H Rugh\ddag}

\address{\dag\ Max-Planck-Institut f{\"u}r Physik komplexer Systeme,
Bayreuther Strasse 40, Haus 16, 01187 Dresden, Germany}

\address{\ddag\ Department of Mathematics, University of Warwick, 
Coventry, CV4 7AL, England}

\begin{abstract}
We present a straightforward and reliable continuous method for computing
the full or a partial Lyapunov spectrum
 associated
with a dynamical system specified by a set of differential equations.
 We do this by introducing a stability parameter $\beta>0$ and
augmenting the dynamical
system with an orthonormal k-dimensional frame
 and a Lyapunov vector
 such that the frame is continuously
Gram-Schmidt orthonormalized
and {\em at most} linear growth of the dynamical
variables is involved. We prove that the method is
strongly stable when $\beta > -\lambda_k$ where $\lambda_k$ is the
$k$'th Lyapunov exponent in descending order
and we show through examples how the method
is implemented. It extends many previous results.
\end{abstract}

\pacs {05.45}

\maketitle

\def\sepand{\rule{14cm}{0pt}\and}
\newcommand{\beq}{\begin{equation}}
\newcommand{\eeq}{\end{equation}}
\newcommand{\bq}{\begin{quotation}}
\newcommand{\eq}{\end{quotation}}
\newcommand{\bc}{\begin{center}}
\newcommand{\ec}{\end{center}}
\newcommand{\BFACE}[1] {\mbox{\boldmath $#1$} }
\newcommand{\Plaq} {{\cal P}}
\newcommand{\half} {\frac{1}{2}}
\newcommand{\r}    {\rightarrow}

\section{Introduction}
One of the most striking coordinate independent
characterizations of a compact dynamical system of dimension $d$
is its Lyapunov spectrum.
It associates to each orbit of the system
a set of $d$ 
real values
which describes
 exponential instabilities of infinitesimal deviations
from the orbit.
Furthermore, for an ergodic dynamical system
this set almost surely (in the measure theoretical sense) does not depend
on which orbit you consider. 
More precisely, given a smooth vector field $v$ in $R^d$ we look
at the $d$-dimensional
 evolution equation $\dot{x} = v(x)$.
For an initial
condition $x(0)=x_0$ we integrate this 
 to obtain a corresponding orbit $x(t) = \phi^t(x_0)$.
The stability of such an orbit can then be examined
by looking at the evolution of a nearby orbit $x(t)+u(t)$
and linearizing the equations of motion in $u$~:
\beq
 \dot{u}(t) = \frac{\partial v}{\partial x}(x(t)) u(t)
      \equiv J(x(t)) u(t)  .
 \label{eq:linear}
\eeq
Integrating this along the orbit we obtain the tangent map
$u(t) = M_{x_0}(t) u_0$ in which the transition matrix
 $M_{x_0}(t) = \partial \phi^t(x_0) / \partial x_0$ is a
$d$ by $d$ matrix valued function of time.
The exponential instabilities of a trajectory
are now reflected
in its eigenvalue spectrum or rather 
the spectrum of the symmetric product
\beq
 M_{x_0}^T(t) M_{x_0}(t)  . \label{eq:symmat}
\eeq
The spectrum of this matrix is real and positive and we order it
as follows~:
\beq
 \mu_1^2(t) \geq \mu_2^2(t) \geq ... \geq \mu_d^2(t) 
 \geq 0 .
  \label {eq:Spectrum}
\eeq
Although the values a priori depends on the initial point $x_0$ chosen,
one has~:\\

\underline{Theorem (Oseledec, \cite{Oseledec,Ruelle})} :
{\em If $\mu$ is an ergodic probability measure for the
dynamical system  then for $\mu$-almost every
$x_0$~:
 \beq \lambda_k = \lim_{t\r\infty} \frac{1}{t} \log \mu_k(t) 
 \label {eq:spectrum}
 \eeq
exists and is independent of the initial point.}\\

In other words taking an arbitrary (with respect to an ergodic
measure) initial point and calculating the above limits, with
probability one you will get its unique Lyapunov spectrum,
$\{\lambda_1 \geq \lambda_2 \geq ... \geq \lambda_d\}$.
  From its definition it is not difficult to show
that the spectrum is independent of the choice of
coordinate system and thus being intrinsic invariants,
their existence and the above theorem is of fundamental
importance in the theory of dynamical systems. Determining
the spectrum (or part of it) has in fact grown into a small
industry in modern non-linear physics.

  From the practical or numerical  point of view the above description
is insufficient as the 
matrix $M^T(t) M(t)$ pretty fast becomes singular
 since its eigenvalues separate
exponentially in time (assuming that not all Lyapunov exponents are equal)
thus making it difficult to measure the spectrum.
Now, several methods have been developed
in order to overcome this problem.
 Fr\o yland \cite{Froyland} uses a
systematic but rather complicated evolution
equation for co-matrices which still has
exponential growth, although in a controllable
way. Meyer \cite{Meyer} makes use of symplectic transformations
to derive the spectrum for Hamiltonian systems also involving
exponential growth. Habib et al. \cite{Habib} uses a 
particular representation of
the Lie algebra so(2) to obtain an evolution
equation for a Hamiltonian system in 2 dimensions
(which appears to work also in dimension 4)  which only
involves linear growth but which is rather complicated and 
depends on the Hamiltonian form of the dynamical system.
Bennetin et al. \cite{Benettin} (for one exponent)
 and later Shimada et al. \cite{Shimada} (for the whole spectrum) suggest
to renormalize (Gram-Schmidt)
 at regular intervals of time a set of stability vectors
picking up the exponents during the renormalization procedure.
This method works well and is often used
 in practice when calculating the spectrum (an example is Gong
\cite{Gong} in which the authors considers the Lyapunov spectrum
for a compact lattice Yang-Mills SU(2) theory). 
Goldhirsch et al. 
\cite{Goldhirsch} present a continuous version of this procedure
(cf. below)
 and they develop a set
of differential equations for the eigenvalues and
eigenvectors of the stability matrix $M_{x_0}(t)$ itself,
a method, however,  unsuited in presence of a  degeneracy of eigenvalues.

Here we shall present a unified approach in which we augment
the dynamical system with an orthonormal frame and a
Lyapunov vector such that the augmented system is 
dynamically strongly stable and involves 
at most linear growth and such that
 the Lyapunov spectrum is
obtained almost surely (in the measure sense of
choosing an arbitrary initial point and frame).
The method is not constrained to Hamiltonian systems,
it applies to any finite dimensional dynamical system
 and is straightforward to implement on a computer. \\

\section{Continuous Gram-Schmidt orthonormalization}

We define a time dependent 
 orthonormal $k$-frame to be a set of $k$
$(k \leq d)$ orthonormal vectors~:
\begin{equation}
 {\cal E}(t) = \{e_1(t),..,e_k(t)\}, \ \
    (e_i,e_j) \equiv \delta_{ij},\ \ 1 \leq i,j \leq k  ,
\end{equation}
where $(\cdot,\cdot)$ is the usual Euclidian product in $R^d$.
Using this frame we let
$J_{lm} = (e_l, J e_m)$ denote the matrix elements of
the Jacobian matrix  $J$ 
and we note that these matrix elements depend on time both through
the Jacobian and the frame. We introduce a stability parameter
 $\beta >0$ 
and define the (symmetric) stabilized
matrix elements
$L_{mm} = J_{mm} + \beta ((e_m,e_m)-1)$ and
$L_{lm} = J_{lm}+J_{ml} + 2 \beta (e_l,e_m)$.
Finally,
let $\Lambda = \{\Lambda_1(t),..., \Lambda_k(t)\}$
be a $k$-dimensional real vector.\\

The augmented dynamical system is now given by the
following set of
differential equations (of which the first two are vector equations)~:

\beq
 \begin{array}{lcll}
   \dot{x} &=& v(x) \ ,& \\
   \dot{e}_m &=& J e_m -  \sum_{l \leq m} e_l L_{lm}
                    & \ \ m=1,...,k \ ,\\
   \dot{\Lambda}_m &=& J_{mm}
                    & \ \ m=1,...,k \ .\\
 \end{array}
 \label{eq:augment}
\eeq

\noindent
For the
dynamical evolution of these equations we have~:\\

\noindent \underline{Theorem} :
 {\em
 Let $x_0$ be an initial point for which 
the associated Lyapunov spectrum 
(cf. equation~\ref{eq:spectrum})
$\lambda_1 \geq \lambda_2 \geq ... \geq \lambda_d$ 
exists. Set $\Lambda(t=0) = 0$.
 Choosing the stability parameter
 $\beta > -\lambda_k$ then for 
almost any (i.e. with probability 1
when choosing randomly) initial frame ${\cal E}(t=0)$
the time evolution of the dynamical system
(\ref{eq:augment}) yields~:
\begin{equation} 
\lim_{t \r \infty} \frac {1}{t} \Lambda_m (t) = \lambda_m  ,
   \ \ m=1,...,k . 
\end{equation}
}

Thus, by following a trajectory of the augmented system
we obtain almost surely
the $k$ first  elements in the Lyapunov spectrum for the given
orbit. The somewhat peculiar condition on the stability parameter
is satisfied e.g. by setting $\beta > \max_{\|e\|=1} (-(e,Je))$
where the maximum is over all unit-length vectors $e$ and over
the relevant region of phase space.
Dynamically such a choice
corresponds to finding the strongest local contraction.

The proof of the Theorem is given in appendix A where in particular, it
is shown that 
the dynamics preserve orthonormality of the frame.
When the elements of $J$ are assumed bounded in phase space
we see that the above dynamical system only involves at most linear
growth of the dynamical variables (through the $\Lambda_m$'s).
   Given a dynamical system with an ergodic measure
we see by combining the Oseledec Theorem and the above that
the Lyapunov spectrum is obtained almost surely
by choosing an arbitrary initial point and an arbitrary initial frame.

An interesting case is when $k=d$, i.e. when we want to calculate
the complete spectrum. In this case, our orthonormal frame is complete
and we may expand $J e_m$ in equation (\ref{eq:augment})
in terms of the basis vectors themselves. Setting $\beta=0$
we get 
\beq
\dot{e}_m = \sum_{l > m} e_l J_{lm} - \sum_{l<m} e_l J_{ml}
  \equiv \sum_{l} e_l A_{lm} ,
\eeq
 where $A_{lm}$ is anti-symmetric
and thus by construction a generator of orthonormal transformations
(of our frame). A straight-forward linear analysis shows that the
resulting dynamical system is marginally stable.\\

\section{Numerical results}
In the following we apply the above method in order to calculate the
Lyapunov spectrum for two standard systems; the
Lorentz system and a 3 degrees of freedom Hamiltonian system with a
quartic potential. In order
to get good statistics on the Lyapunov exponents we determine a time
over which to integrate the systems in order to get reasonable
convergence and then make 1000 runs of each system with random initial 
conditions.

\begin{figure}
\centerline{\epsfig{file=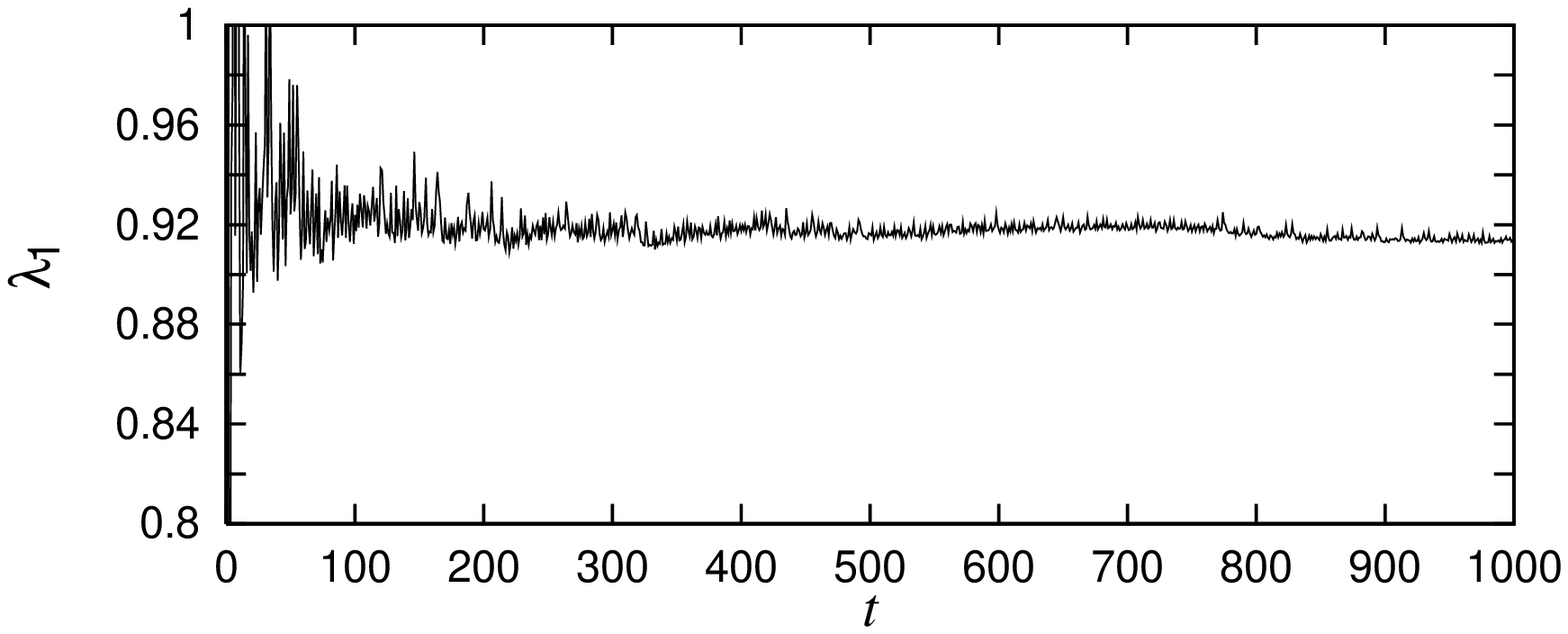,width=12cm}}
\centerline{\epsfig{file=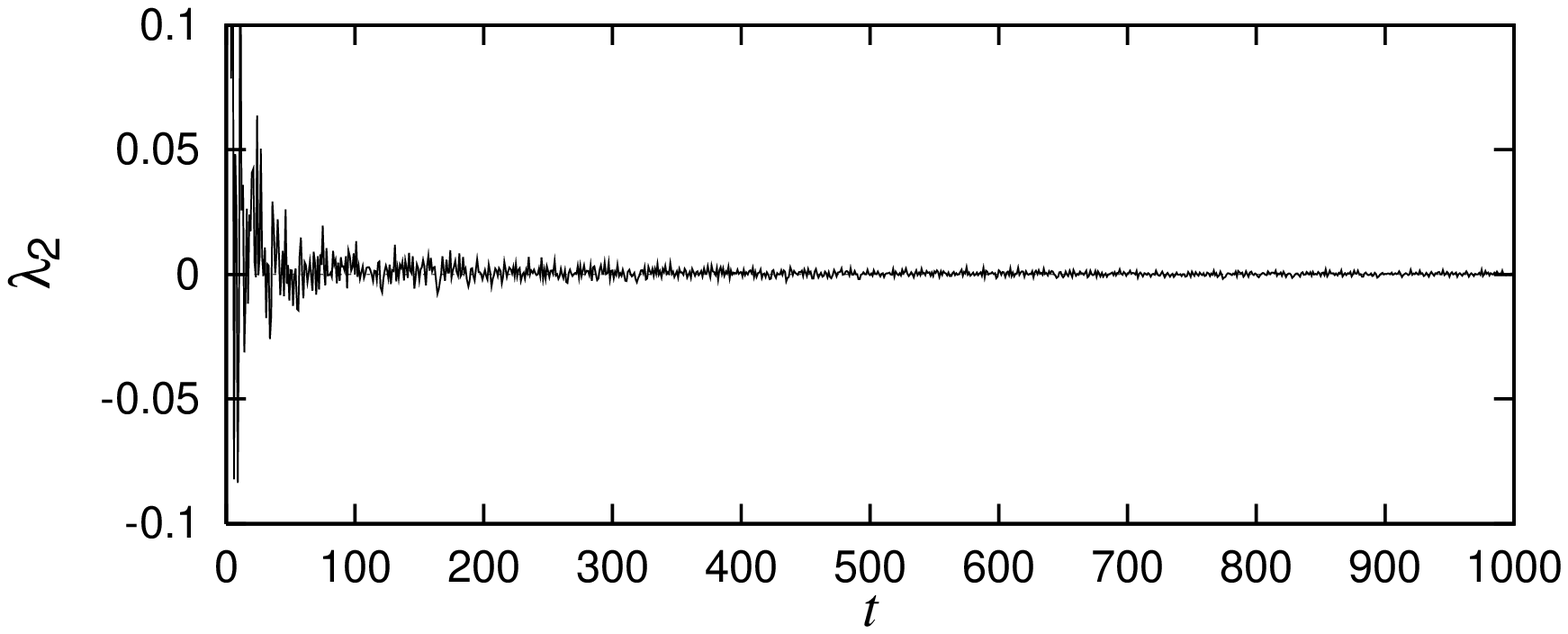,width=12cm}}
\centerline{\epsfig{file=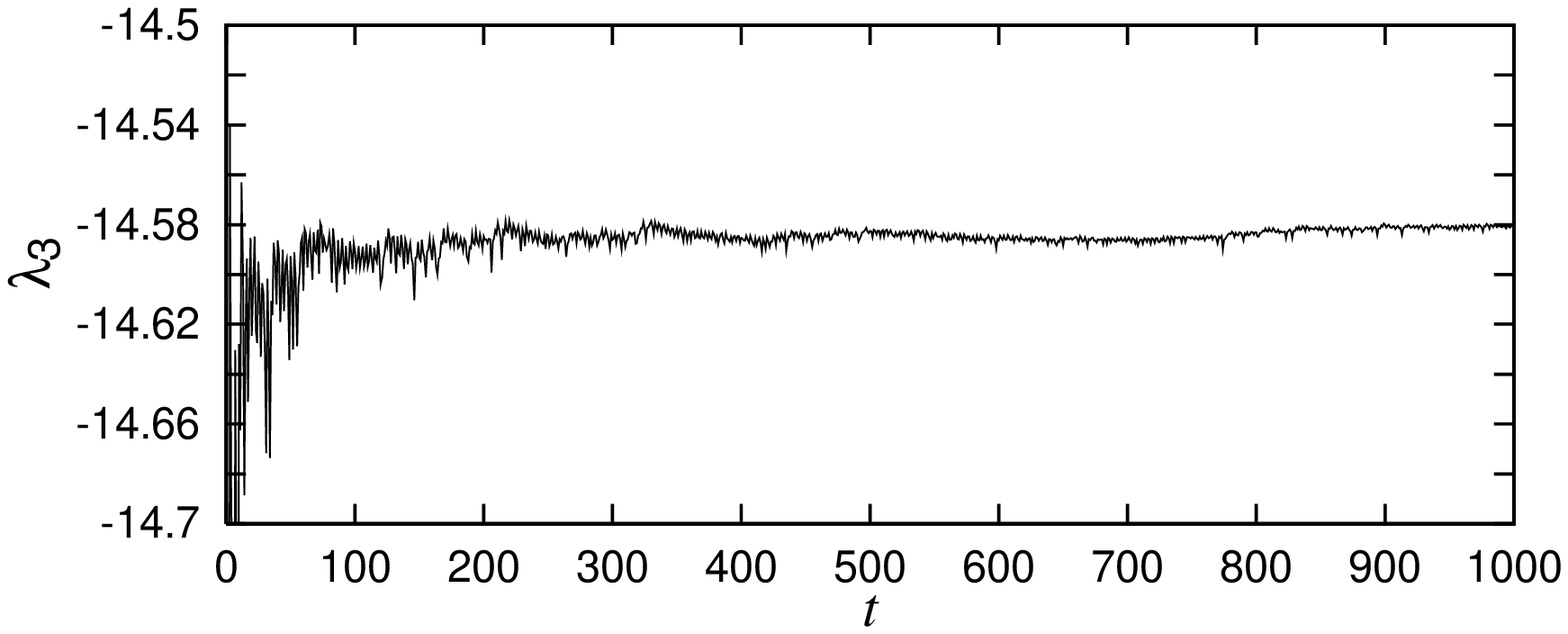,width=12cm}}
\caption[lor]{The 3 (finite-time) Lyapunov exponents from a single run 
of the Lorentz system.}
\label{lorfig}
\end{figure}

The Lorentz system~\cite{lorentz} is given by:
\begin{eqnarray}
\label{eqlor}
\dot{x} & = & -\sigma x+\sigma y   \nonumber \\
\dot{y} & = & r x -y -xz \\
\dot{z} & = & xy -b z  \nonumber
\end{eqnarray}
with the usual choice of parameter $\sigma=10$, $r=28$ and $b=8/3$.
In figure \ref{lorfig} we present the resulting Lyapunov 
exponents of a single run for the Lorentz system. The result of 1000 runs
is given in table \ref{lortab}.

\begin{table}
\caption[captablor]{The average of the Lyapunov exponents of 1000 runs of
the Lorentz system and their root mean square deviations. The sum of the
exponents is $-13.6667=-\sigma-1-b$ as expected from
equation  (\ref{eqlor}).}
\begin{indented}
\lineup
\item[]\begin{tabular}{@{}lll}
\br
$k$ & $\lambda_k$ & rms dev. \\ \mr
1 & \00.9057 & 4.7$\cdot 10^{-3}$ \\
2 & \01.4$\cdot 10^{-5}$ & 8.3$\cdot 10^{-4}$ \\
3 & \-14.5724 & 4.6$\cdot 10^{-3}$ \\ \br
\end{tabular}
\end{indented}
\label{lortab}
\end{table}

\begin{figure}
\centerline{\epsfig{file=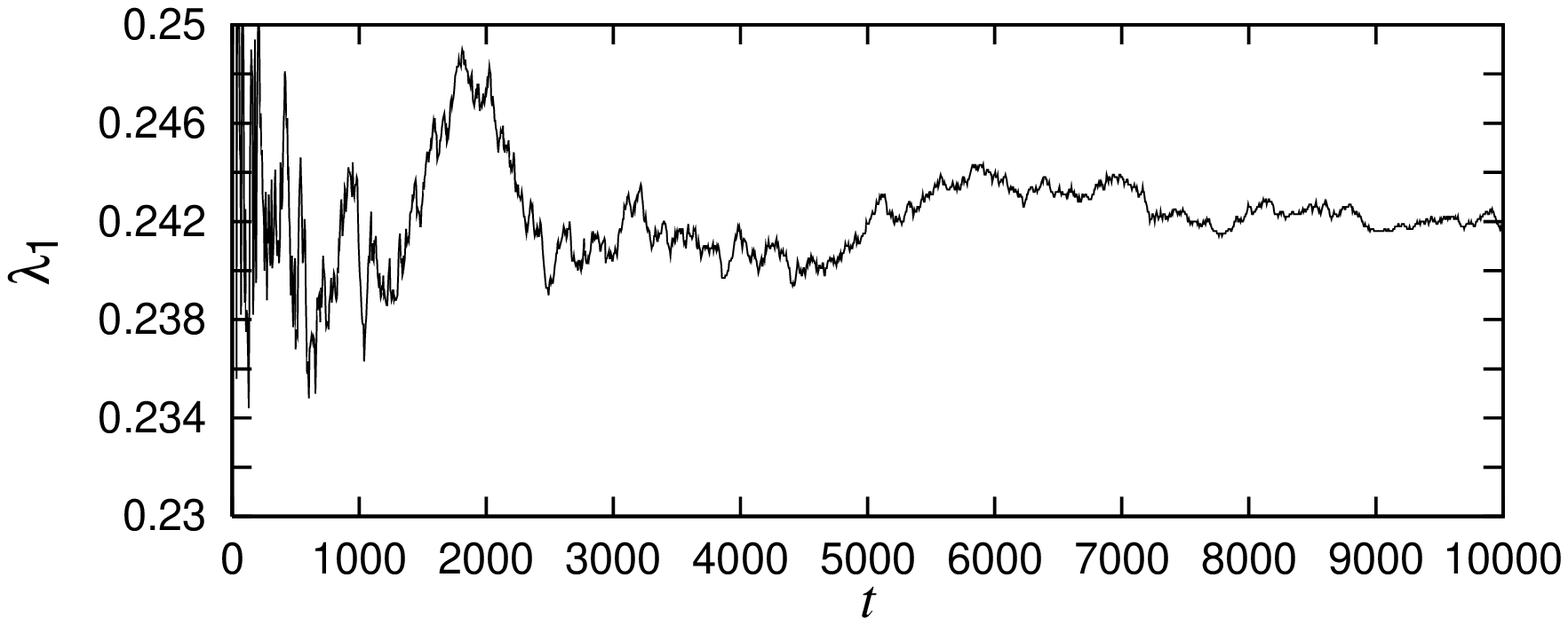,width=12cm}}
\centerline{\epsfig{file=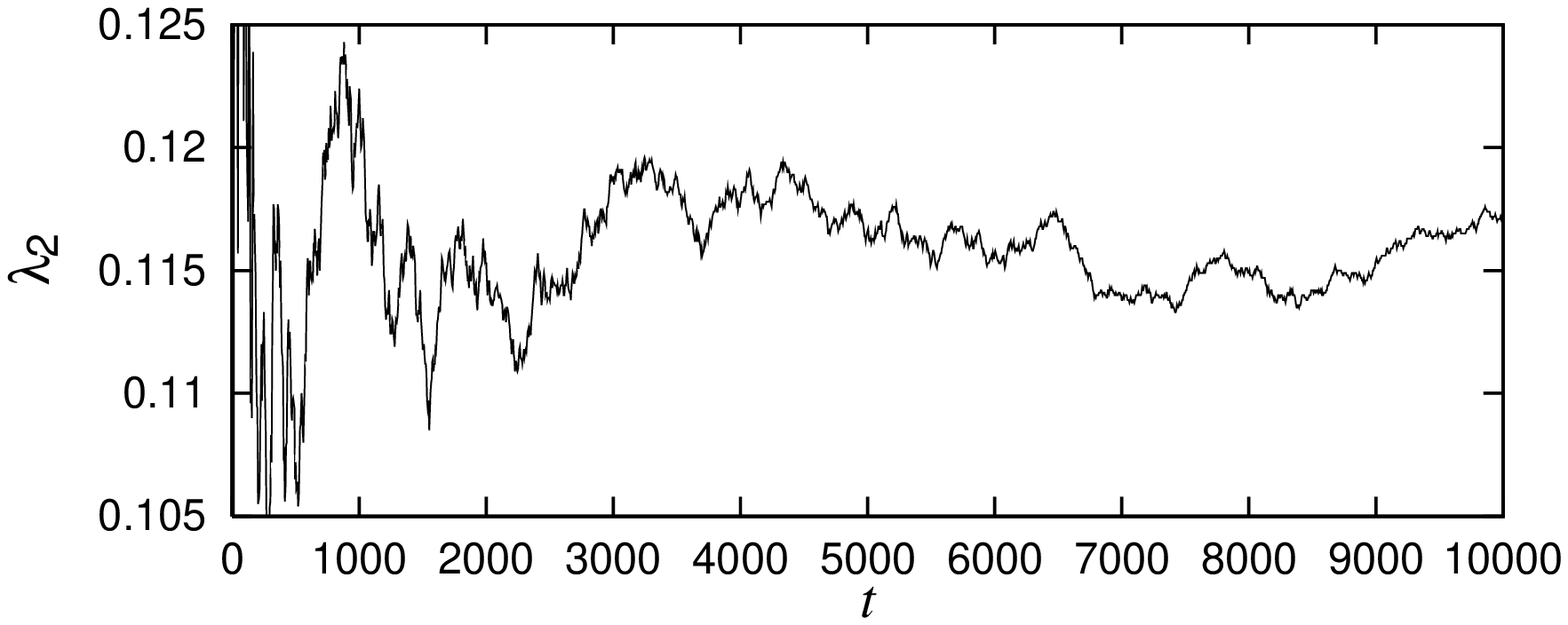,width=12cm}}
\centerline{\epsfig{file=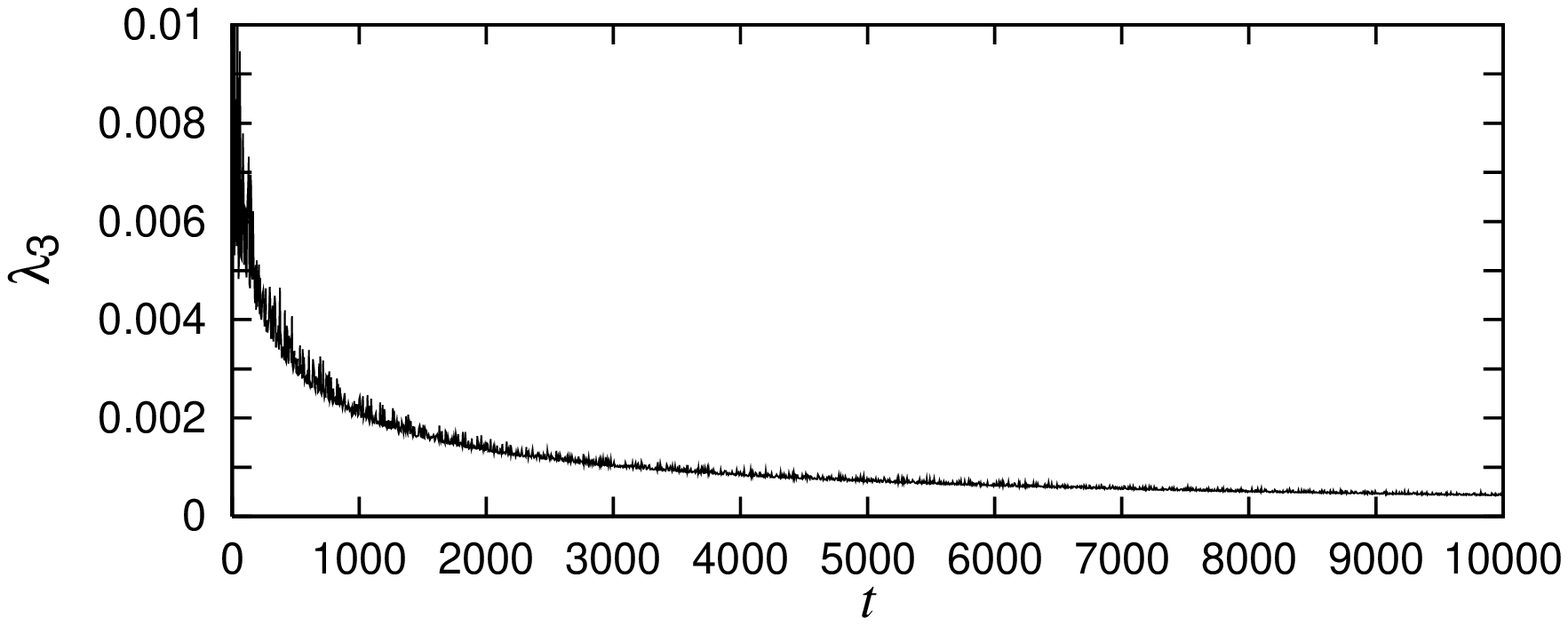,width=12cm}}
\caption[ym]{The 3 positive (finite-time) Lyapunov exponents from a 
single run of (\ref{ymeq}). We present only the
positive exponents since, due to the symplectic structure of the equations,
the absolute value of the negative exponents are exactly equal to 
the positive ones.}
\label{ymfig}
\end{figure}

Next we apply our method to a Hamiltonian system 
with a quartic potential:
\begin{equation}
H= \frac{p_{x}^2 + p_{y}^2 + p_{z}^2}{2} +\frac{x^2y^2+y^2z^2+z^2x^2}{2} +
\frac{x^4+y^4+z^4}{32} .
\label{ymeq}
\end{equation}
The last term is added in order to have a compact phase space and
avoid having to deal with problems of convergence related to near-integrable
motion along
the coordinate axes, but still chosen with a sufficiently small prefactor
in order not to stabilize the dynamics.
A single run of the resulting 6-$d$ system is presented in figure \ref{ymfig}
and the corresponding averages of 1000 runs are given in table \ref{ymtab}.

\begin{table}
\caption[captabym]{The average of the Lyapunov exponents of 1000 runs of
the Hamiltonian system (\ref{ymeq}) and their root mean square deviations.}
\begin{indented}
\lineup
\item[]\begin{tabular}{@{}lll}
\br
$k$ & $\lambda_k$ & rms dev. \\ \mr
1 & 0.2374 & 3.6$\cdot 10^{-3}$ \\
2 & 0.1184 & 3.6$\cdot 10^{-3}$ \\ 
3 & 3.90$\cdot 10^{-4}$ & 7.0$\cdot 10^{-5}$ \\ \br
\end{tabular}
\end{indented}
\label{ymtab}
\end{table}

\begin{figure}
\centerline{\epsfig{file=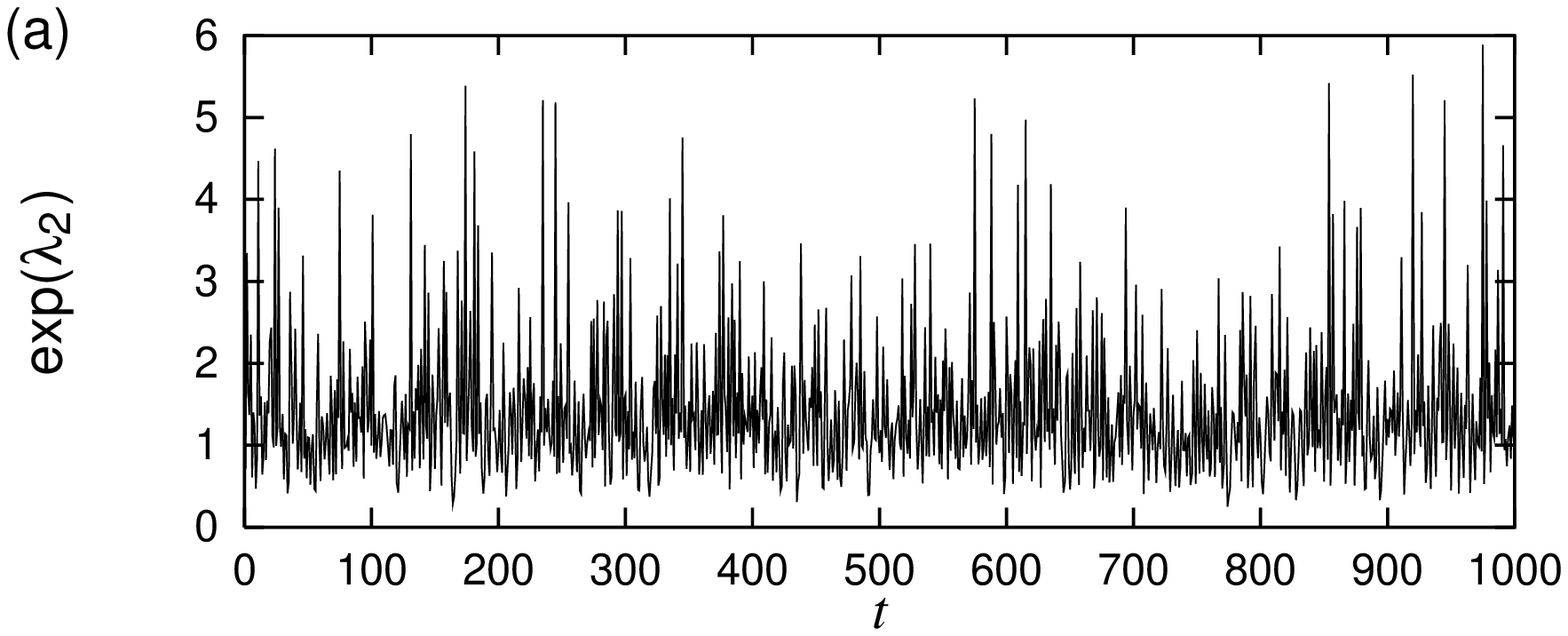,width=12cm}}
\centerline{\epsfig{file=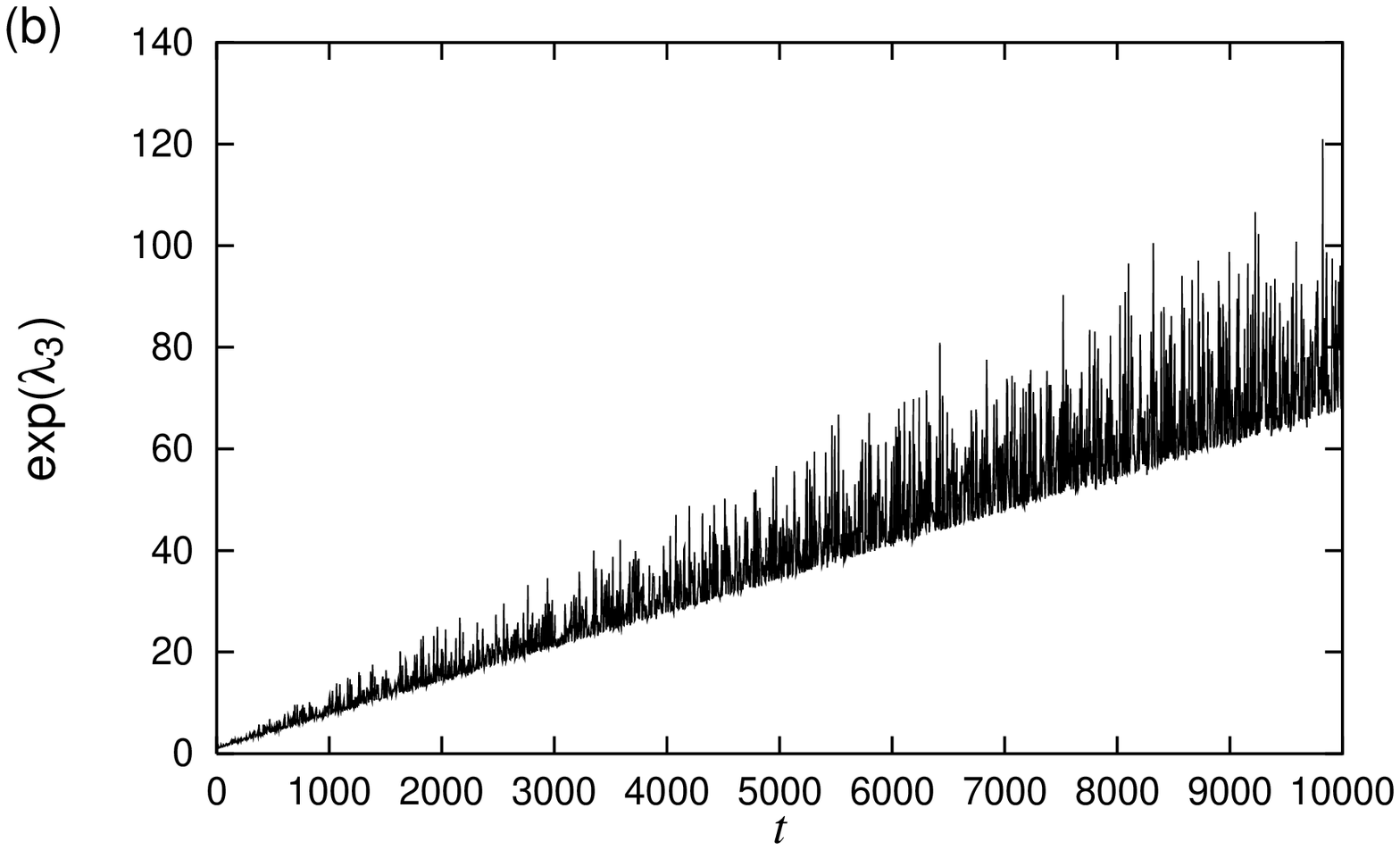,width=12cm}}
\caption[figeig]{$\e^{\lambda t}$ where $\lambda$ is the exponent corresponding
to the marginally stable direction along the flow for (a) the Lorentz system
and (b) the Hamiltonian system (\ref{ymeq}).}
\label{eigfig}
\end{figure}

Looking at the results for the two systems there is one striking
difference which is apparent both for the single run results shown in the
figures and in the averages given in the tables. Whereas the (finite-time)
exponent
$\lambda_2$ for the Lorentz system, corresponding to the marginally stable
direction along the flow, fluctuates around zero, the equivalent (finite-time)
exponent,
$\lambda_3$, for the Hamiltonian system is clearly different from zero though
converging for increasing $t$. The difference can be made even more apparent
when plotting instead $\e^{\lambda t}$ as in figure \ref{eigfig}. 
$\e^{\lambda t}$ is the stability eigenvalue for the marginally stable
direction over the entire integration. For the Lorentz system this remains
constantly close to one as expected by a marginal eigenvalue. For the
Hamiltonian system, however, it grows linearly. The explanation
lies in the degeneracy of this eigenvalue. The Hamiltonian (\ref{ymeq}) has
another marginally stable direction associated with grad$H$. 
In local coordinates the
Jacobian for the flow will in general take on a Jordan normal form for
the two marginally stable directions, i.e. we can not expect to find a full
set of eigenvectors for the Jacobian. One may consider the possibility of
factoring out the marginally stable directions and not include them in the
integration, but here we just note that we know that the corresponding
exponents exactly equals zero for infinite time.

\begin{figure}
\centerline{\epsfig{file=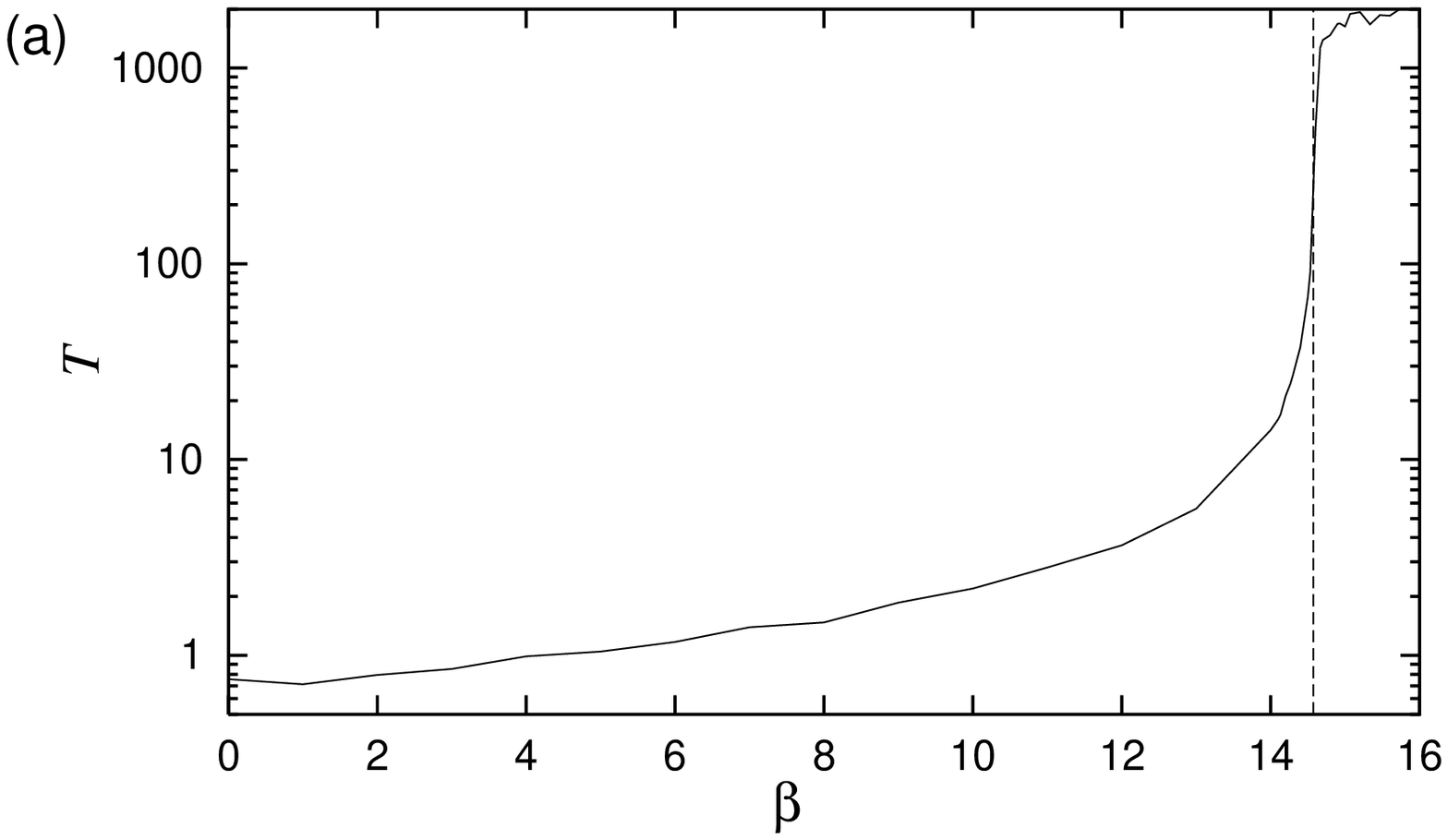,width=12cm}}
\centerline{\epsfig{file=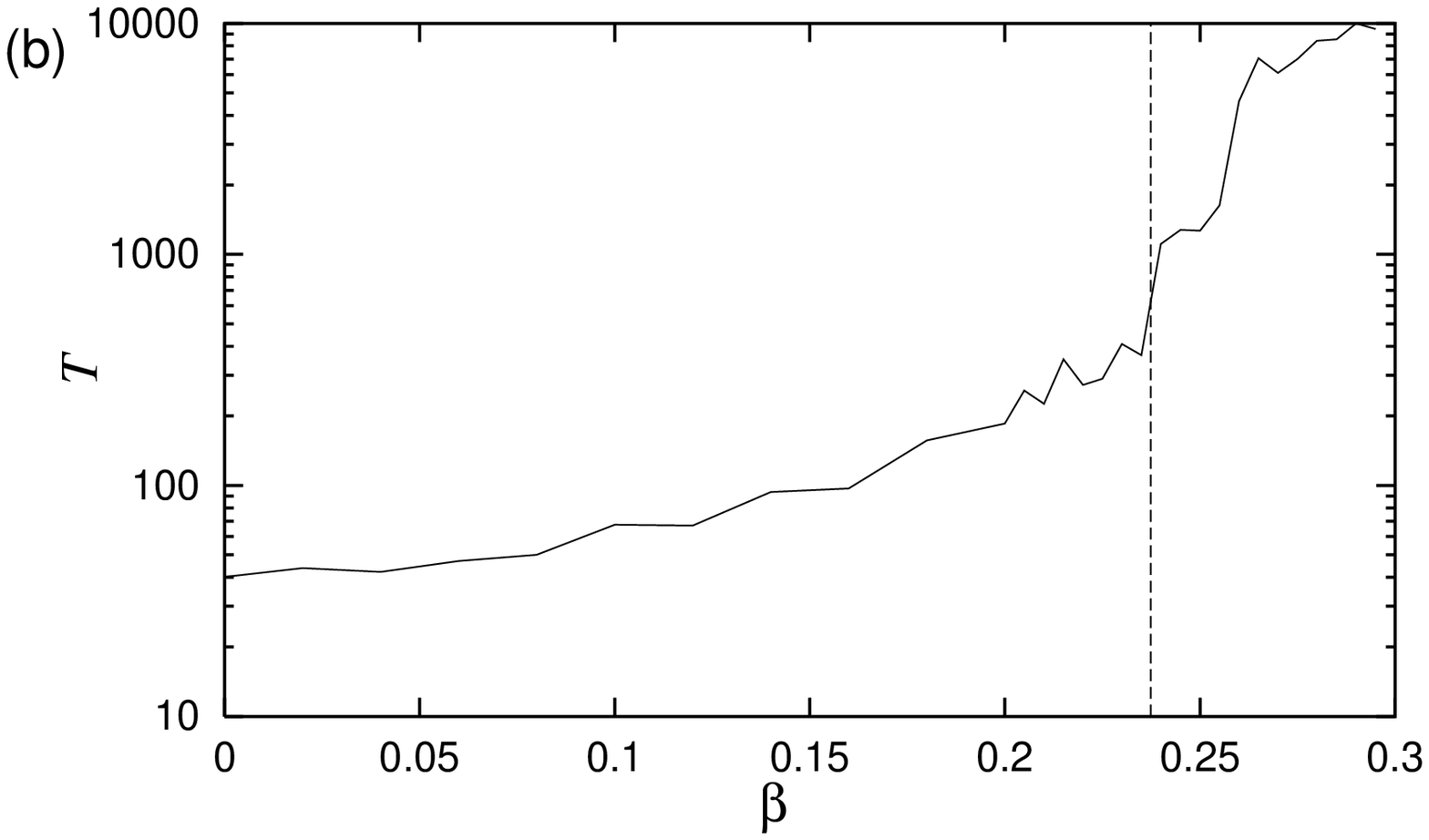,width=12cm}}
\caption[tim]{The time $T$ it takes to reach an error level of $10^{-3}$
on the orthonormality of the basis $\cal{E}$ for the Lorentz system (a) and
the Hamiltonian system (\ref{ymeq}). }
\label{time}
\end{figure}

To illustrate the effect of the stabilizing factor $\beta$ we consider the
time to reach a certain error level on the orthonormality of the basis
$\cal{E}$ with varying $\beta$. To be more specific we consider the
error $e=[\sum_{i,j=1,k} ((e_i,e_j)-\delta_{ij})^2 ]^{1/2}$ and the 
time $T(\beta)$ it
takes for this error to grow to a certain level as a function of $\beta$.
Since this will depend on the initial conditions we have taken the average
of 10 runs for each of the systems. 
The result is presented in figure \ref{time}. 
We let the systems run a maximum
time before deciding that the chosen error level would not be reached. For
the Lorentz system this time was set at 2000 and for the Hamiltonian system
at 10000. The curves of figure \ref{time} therefore saturate at these values.
For the Lorentz system $T(\beta)$ increases drastically near 
$\beta=-\lambda_3$ as expected, whereas for the Hamiltonian system the
picture is a little less clear with the large increase of $T(\beta)$ happening
at a slightly larger value than $\beta=-\lambda_6$. The difference is due to
the relative high homogeneity of the Lorentz system vs. the rather strong
dependence of the local finite-time stability exponents on the phase space
position in the Hamiltonian system. Based on these results we have chosen
to set $\beta=20$ for the Lorentz system and $\beta=.5$ for the Hamiltonian 
system. The point here is to set $\beta$ sufficiently high to stabilize the
given system under integration, but not excessively high since this could
easily lead to unnecessarily high requirements on the accuracy of the 
integration routine.\\

The method to calculate Lyapunov exponents of ODE's that
we have presented in this paper is nothing
but a continuous version of the standard Gram-Schmidt orthonormalization
procedure.  As mentioned above this was already proposed by
Goldhirsch et al.  \cite[eqs. (5.12) and (5.26)]{Goldhirsch} 
but without the crucial stability term. 
 Apart from the aesthetic pleasure of formulating the whole
procedure as one set of differential equations, the method will show its
usefulness when calculating Lyapunov spectra where the difference between
the largest and the smallest exponent is large. In such a case,
using standard methods, one 
rapidly loses accuracy on the eigenvector associated with the lowest exponent
and therefore also of the exponent itself. One would therefore have to employ
the Gram-Schmidt orthonormalization quite often; the continuous
orthonormalization naturally avoids this problem with
the $\beta$-term replacing the stabilizing effect of the (non-linear)
Gram-Schmidt procedure.
 On the other hand, if one
is only interested in calculating the largest exponent for a given system
there is essentially no difference between standard methods and the method
given here since orthonormalization is unnecessary except to avoid a possible
numerical overflow. Computationally the continuous method is somewhat heavier 
than standard methods. To compute $k$ exponents of a $d$-dimensional
system one needs, in addition to the usual $Je_m$ operation (${\cal O}(kd^2)$),
to compute $e_l Je_m$ and $e_l e_m$, both ${\cal O}(k^2 d)$. For a full 
spectrum the computation is thus a factor of 3 heavier, whereas for
partial spectra it will be somewhat less.

\newpage

\appendix
\section{}
\subsection{Proof of the Theorem}

\newcommand{\Llm}{\kappa_{lm}}
\newcommand{\Lln}{\kappa_{ln}}
\newcommand{\dLlm}{\dot{\kappa}_{lm}}
\newcommand{\Lmm}{\kappa_{mm}}
\newcommand{\dLmm}{\dot{\kappa}_{mm}}
\newcommand{\dem}{\dot{e}_{m}}
\newcommand{\del}{\dot{e}_{l}}

We first remark that when the frame is orthonormal the
stabilizing terms vanish identically. We shall show that
the resulting equations correspond to
a differential version
of a Gram-Schmidt orthonormalization of a
set of $k$ independent   tangent vectors
evolving in time according to 
equation (\ref{eq:linear}).

Thus, consider 
a time dependent set of vectors $\{f_1,...,f_k\}$ which are
linearly independent and satisfies
$\dot{f}_m = J(x(t)) f_m$. We expand this set of vectors
uniquely  into an orthonormal set
$\{e_1,...,e_k\}$, i.e.
$ f_m = \sum_{l \leq m} e_l \Llm$ where
$\{\kappa_{lm}\}_{l \leq m}$ are a set of time dependent
parameters with positive diagonal elements, i.e. $\kappa_{mm} >0$.

We shall show that the $e_m$'s and the diagonal elements $\Lmm$ 
satisfy the differential equations~:

\beq
 \begin{array}{lll}
 \dot{e}_m  &= & J e_m - e_m J_{mm} - \sum_{l < m} e_l (J_{lm}+J_{ml}) \ , \\
 \dLmm &=& J_{mm} \Lmm \ .
 \label{eq:gs}
 \end{array}
\eeq

We prove it by induction. So assume it is true
for the vectors $\{e_1,...,e_{m-1}\}$.
  From $\dot{f}_m = J f_m$ we get~:
\beq
 \sum_{l \leq m} (\del \Llm + e_l \dLlm - J e_l \Llm) = 0 \ .
 \label{eq:eq}
\eeq
By orthonormality $(e_m,\dem)=0$ and by the induction hypothesis
$(e_m,\del) = J_{ml}$ for $l<m$ so we obtain by taking
the scalar product with $e_m$~:
\beq 
\sum_{l<m} (J_{ml} \Llm) + \dLmm - \sum_{l\leq m} J_{ml} \Llm = 0
\eeq
and hence that $\dLmm = J_{mm} \Lmm$. Again for $l<m$ by the induction
hypothesis $\del - J e_l$ and hence also
$ \sum_{l<m} (\del \Llm + e_l \dLlm - J e_l \Llm)$ 
 are  in the span of $e_1,...,e_{m-1}$
(just re-arrange the terms in (\ref{eq:gs})) and hence, we may
rewrite equation (\ref{eq:eq}) as follows~:
\beq
 \dem \Lmm + e_m J_{mm} \Lmm - J e_m \Lmm = \sum_{l < m} e_l c_l \ , 
 \label{eq:eqq}
\eeq
where $c_l$ are some (time dependent) parameters. 
For $l<m$ the scalar
product $(e_l,e_m)$ is constant (zero) in time.
Whence $(e_l,\dem) = - (e_m,\del) = -J_{ml}$ and
we may take the
scalar product with $e_l$ in (\ref{eq:eqq})
to obtain
\beq
 -(J_{ml} + J_{lm}) \Lmm = c_l \ .
\eeq
Finally, inserting this in (\ref{eq:eqq})
and dividing by $\Lmm$
we get the desired equation for $\dot{e}_m$.
  From (\ref{eq:augment}) and the above we get the relationship
$\Lmm(t) = \exp (\Lambda_m(t)) \Lmm(0)$.\\

\subsection{The Lyapunov spectrum}
Next, we will show that provided the Lyapunov spectrum exists
(we assume so from now on),
the limits $\lim_{t \r \infty} \frac{1}{t} \Lambda_m(t)$ will
almost surely give the
spectrum (in descending order). In order to do this we take
the positive matrix $M^T M$ from equation (\ref{eq:symmat}) and
diagonalize it to obtain~:
\beq
   M^T M = \sum_m \mu_m^2 a_m \otimes a_m 
\label{eq:diag}
\eeq
where $\{a_k\}_{m=1,..,k}$ is a set of orthonormal vectors
and the $\mu_k$'s are as in (\ref{eq:Spectrum}).
One has $\| Mu \|^2 = \sum \mu_k^2 (a_k,u)^2$
which geometrically means that the image of a sphere
$\|u\|=1$ will be an ellipsoidal with principal
axes $\{\mu_k\}$. Now let $1>r>0$ be a fixed constant
and let $u$ be a unit length vector.
 Suppose that $|(a_1,u)| > r >0$
at all times.
Then we have~:
\beq
 \mu_1^2 \geq \|Mu\|^2 \geq r^2 \mu_1^2 
\eeq
and hence it follows  that~:
\beq
  \lim_{t\r \infty} \frac{1}{2t} \log \|Mu\|^2 - \lambda_1
 = \lim_{t\r \infty} \frac{1}{t} \log \mu_1  - \lambda_1
  + \frac{1}{t} (\log \|Mu\| - \log \mu_1) = 0 
\eeq
since
$\lambda_1 = \lim \frac{1}{t} \log \mu_1$ and
 the expression in the parenthesis is uniformly bounded~:
$\log(r) \leq \log \|Mu\| - \log \mu_1 \leq 0$.
Now, the vector $a_1$ actually depends on time, but 
since $u$ is chosen at random we have at any given instant 
that $|(a_1,u)| > r >0$ with a probability $p(r)$ which
tends to 1 as $r$ tends to zero. This follows from simple
geometrical considerations on the area of the $d$-ball,
compared to the part of it for which the above inequality
holds. Hence the above results holds with probability $p(r)$
and since $r>0$ was arbitrary it follows 
that with probability 1 the limit of $1/t \log \|Mu\|$
will exist and be the maximal Lyapunov exponent.
 Inserting here $u=e_1(0)$
and $Mu = \exp(\Lambda_{1}(t)) e_1(t)$ we obtain the
desired result.\\

The general case is shown by considering growth rates
of exterior products.
We shall only show the explicit calculations for the
first two Lyapunov exponents, noting that the formulae easily
generalize.

 We recall that if $\{e_1,...,e_d\}$ is a
basis for $V$, then the formal exterior products
$\{e_k \wedge e_l\}_{k<l}$ is a basis for the
vector space  $\wedge^2 V = V \wedge V$.
One may define the scalar product~:
\beq (u_1\wedge u_2, v_1 \wedge v_2) = \det \left|
    \begin{array}{cc}
         (u_1,v_1) & (u_1,v_2) \\
         (u_2,v_1) & (u_2,v_2) 
    \end{array} \right| \ ,
\eeq
as well as the action of $\wedge^2 A = A \wedge A$~:
\beq 
A\wedge A (u_1 \wedge u_2) = (A u_1) \wedge (A u_2)  \ .
\eeq
Consider now the action of the matrix $\wedge^2 (M^T M)$
in the following way~:
\beq
 \begin{array}{l}
  (u_1 \wedge u_2, \wedge^2 (M^T M) v_1 \wedge v_2) =\\
  (u_1 \wedge u_2, (M^T M v_1) \wedge (M^T M v_2)) =\\
  \det \{ (u_i, M^T M v_j)\} =\\
  \det \{ ( M u_i, M v_j) \} = \\
  (\wedge^2 M  u_1\wedge u_2, \wedge^2 M v_1 \wedge v_2) = \\
  (u_1 \wedge u_2, (\wedge^2 M)^T (\wedge^2 M) v_1 \wedge v_2) .\\
 \end{array}
  \label{eq:appmm}
\eeq
In particular, this shows that
 $\wedge^2 (M^T M) = (\wedge^2 M)^T (\wedge^2 M)$, an identity
which will allow us to estimate the growth rate of the product
of the two largest eigenvalues of $M$. Using the diagonalization
(\ref{eq:diag}) as well as linearity and anti-symmetry of the
wedge product we get~:
\beq 
 \wedge^2 (M^T M) (u_1 \wedge u_2) =
   \sum_{i<j} \mu_i^2 \mu_j^2 a_i\wedge a_j
     (a_i\wedge a_j, u_1\wedge u_2) \ .
\eeq
In particular~:
\beq
 (u_1 \wedge u_2, \wedge^2 (M^T M) (u_1 \wedge u_2)) =
   \sum_{i<j} \mu_i^2 \mu_j^2
     (a_i\wedge a_j, u_1\wedge u_2)^2 \ .
\eeq
and using (\ref{eq:appmm}) we deduce the inequality~:\
 \beq 
\mu_1 \mu_2 |(a_1\wedge a_2, u_1\wedge u_2)|
   \leq \| \wedge^2 M (u_1 \wedge u_2)\| \leq \mu_1 \mu_2 \ .
\eeq

We can then repeat the arguments from above to show that with
probability 1~:
\beq 
\lim \frac{1}{t}  
   \log \| \wedge^2 M (u_1 \wedge u_2)\| = \lambda_1 + \lambda_2 
\eeq
and by anti-symmetry of the wedge 
the left hand side will apart from a uniformly
bounded contribution (which almost surely
 vanishes in the limit) equal 
\beq 
\lim \frac{1}{t} \log \| f_1 \wedge f_2 \| =
    \lim \frac{1}{t} \log |\kappa_{11}(t) \kappa_{22}(t)|
   = \lim \frac{1}{t} (\Lambda_1(t) + \Lambda_2(t)) , 
\eeq
and using the already obtained formula for the exponent
$\lambda_1$ the desired result follows for $\lambda_2$.\\

\subsection{Linear stability theory}
The results obtained above are numerically reliable
provided the frame stays orthonormal during the time evolution.
In particular 
the variables $\Delta_{lm} = (e_l,e_m) - \delta_{lm}$, 
$1 \leq l,m \leq k$ should all vanish.
By straight-forward differentiation one verifies
that they satisfy the following set of differential equations~:
\beq 
\dot{\Delta}_{pm} = -(2 \beta + L_{mm}+L_{pp})
      - \sum_{l<m} \Delta_{pl}L_{lm}
      - \sum_{l<p} \Delta_{lm} L_{lp} .
\eeq
It is clear that $\Delta_{pm} \equiv 0$, 
for all $p$ and $m$, is a fixed point of
these equations. In order to analyze its stability we
substitute $\Delta_{pm} \r \Delta_{pm} + \delta_{pm}$ and
linearize in the variables $\delta_{pm}$
to find~:
\beq 
\dot{\delta}_{pm} = - (2 \beta + J_{mm} + J_{pp}) \delta_{pm}
          + G(\{\delta_{pl}\}_{l<m}, \{\delta_{lm}\}_{l<p}) 
\eeq
where $G$ is linear in the $\delta$'s but depends
{\em only on the preceding variables}.  Here we have
used the natural
lexicographic ordering :
 $(11) < (21)=(12) < (22) < (31)=(13) < (32)=(23) < ...$
It follows that the stability of these equations is 
determined only by the stability of the
first term, i.e. of the differential equations
(for all $p$ and $m$)~:
\beq
 \dot{z} = - (2\beta + J_{mm} + J_{pp}) z .
\label{eq:suf}
\eeq
This equation happens to be analytically solvable, surprisingly in terms
of our Lyapunov vectors themselves,
\beq
 z(t) = \exp (- 2 \beta t - \Lambda_{mm}(t) - \Lambda_{pp}(t)) ,
\eeq
and we see that stability is ensured provided
\beq 
\beta > - \lim_{t \r \infty} \frac{1}{t} \Lambda_{mm}(t)
       = - \lambda_m , 
\eeq
for all $m = 1,...,k$. As our $\lambda$'s are ordered
decreasingly it suffices to have $\beta > -\lambda_k$.
  From equation (\ref{eq:suf}) we also see that
stability follows by using the more
conservative bound obtained by setting
$\beta > \max_{\|e\|=1} (-(e,Je))$.

\end{document}